# Developing effective electronic-only coupled-cluster and Møller-Plesset perturbation theories for the muonic molecules

Mohammad Goli[1,*] and Shant Shahbazian[2,*]

[1] *School of Nano Science, Institute for Research in Fundamental Sciences (IPM), Tehran 19395-5531, Iran*

[2] *Department of Physics, Shahid Beheshti University, G. C., Evin, Tehran, Iran, 19839, P.O. Box 19395-4716.*

E-mails:

Mohammad Goli: m_goli@ipm.ir

Shant Shahbazian: sh_shahbazian@sbu.ac.ir

[*] Corresponding authors

# Abstract

Recently we have proposed an effective Hartree-Fock (EHF) theory for the electrons of the muonic molecules that is formally equivalent to the HF theory within the context of the Nuclear-Electronic Orbital theory [Phys. Chem. Chem. Phys. 20, 4466 (2018)]. In the present report we extend the muon-specific effective electronic structure theory beyond the EHF level by introducing the effective second order Møller-Plesset perturbation theory (EMP2) and the effective coupled-cluster theory at single and double excitation levels (ECCSD) as well as an improved version including perturbative triple excitations (ECCSD(T)). These theories incorporate electron-electron correlation into the effective paradigm and through their computational implementation, a diverse set of small muonic species is considered as a benchmark at these post-EHF levels. A comparative computational study on this set demonstrates that the muonic bond length is in general non-negligibly longer than corresponding hydrogenic analogs. Next, the developed post-EHF theories are applied for the muoniated N-Heterocyclic carbene/silylene/germylene and the muoniated triazolium cation revealing the relative stability of the sticking sites of the muon in each species. The computational results, in line with previously reported experimental data demonstrate that the muon generally prefers to attach to the divalent atom with carbeneic nature. A detailed comparison of these muonic adducts with the corresponding hydrogenic adducts reveals subtle differences that have already been overlooked.

# I. Introduction

Implanting the positively charged muon directly or in the form of the muonium atom, i.e. a muon plus an electron, in molecules is currently a well-established approach to probe local traits of atomic environments in chemistry and solid-state physics.[1-15] Practically, detection of the ejecting positrons, arising from decaying spin-polarized muons, forms the basis of the muon spin resonance spectroscopy to deduce the sticking site of the muon or muonium atom.[16,17] In most cases there are various atoms or bonds in a molecule that may act as potential binding sites making the interpretation of the spectrum non-trivial.[18-23] Thus, it is vital to have a clear picture of the conceivable attachment sites in order to deduce the local electronic information by the spectrum analysis. Without having any other complementary experimental sources on the molecular structure of the muonic species, the only available means to discern the addition sites is theoretical and computational modeling. Various approaches have been proposed to explore the potential attachment sites of the muon or the muonium atom to molecular species.[24-42] One possible route is trying to solve time-independent Schrödinger equation for the muonic molecules, i.e. a molecule plus the muon (μ) or muonium atom (Mu), using the well-established ab initio procedures developed for conventional molecules.[43,44] Nevertheless, the usual adiabatic framework that decomposes molecule into fast and slow moving particles, i.e. assuming nuclei as clamped point charges and electrons as quantum particles at first step, is not a safe framework to be used for the muonic species. This stems from the fact that muon's mass is only 207 times of the electron's mass, not large enough to justify the adiabatic view. In order to circumvent this problem, it seems reasonable to include the muon as a quantum particle from the outset, i.e. to solve the time-independent Schrödinger equation including the kinetic energy operator of the muon. This approach is theoretically straightforward since in the last twenty years various ab initio theories have been developed to extend the quantitative molecular orbital method to the *multi-component* quantum systems treating both electrons and nuclei as quantum particles.[45-49] In this framework, the muonic species are two-component quantum systems composed of electrons and the muon within the external electric field generated by the clamped nuclei. Recently, using the Nuclear-Electronic Orbital (NEO) non-adiabatic ab initio theory,[46] the Nuclear-Electronic Orbital Hartree-Fock (NEO-HF) equations have been implemented computationally for the muonic species.[50-53] The NEO-HF equations extend the molecular orbital method to the muonic species attributing orbitals not only to electrons but also to the muon even though the muonic orbital describes the vibrations of the muon. In the next step the electron-electron (ee) and muon-electron (μe) correlations must be incorporated into the post-NEO-HF equations as implemented originally in the NEO theory.[46,54] This scheme is, in principle, fully extendable to the muonic species reaching eventually to the full configuration interaction (NEO-FCI) method as the exact solution of Schrodinger's equation within the given basis sets.[46]

Instead of applying the NEO-HF theory directly to the muonic species, we have recently proposed an alternative formalism, which is a reformulation of the NEO-HF theory, and is called the effective HF (EHF) theory.[55,56] Within the context of the EHF theory, the muon disappears from the molecular Hamiltonian with the aid of integration over the muonic coordinates, and a as result, a non-Coulombic potential complements the effective electronic Fock operator. This additional term would be optimized during the self-consistent field solution of the EHF equations,[56] and is directly derivable from the basis set used to expand the muonic orbital manifesting the "back-action" of muon's vibrations on the electronic distribution; after solving the EHF equations, the muonic orbital is completely reproducible from the optimized effective potential. It is timely to stress that we have recently extended the idea of the effective theory to the NEO density functional theory (NEO-DFT) and derived an effective set of electronic-only Kohn-Sham (EKS) equations for the muonic molecules.[57]

In this report, the effective theory is extended beyond the mean-field approximation and the *ee* correlation is accounted for using an effective Hamiltonian for the muonic molecules. This is done within the context of the second order Møller-Plesset (MP2) perturbation theory and the coupled-cluster (CC) theory as well-known examples of non-variational methods. The proposed idea is quite general thus can be extended to the configuration interaction and the multi-configurational self-consistent field theories as variational methods. The estimation of the *μe* correlation is a more complex task and in contrast to efforts directed toward proper description of the nucleus-electron correlation,[58-79] no theoretically sound, yet computationally tractable "universal" solution to this problem has been proposed. By the way, it conceivable that certain *ad hoc* adjustments of the effective Hamiltonian introduced in the present study, like the modification proposed by Kerridge and coworkers,[80] may partly incorporate the *μe* correlation into the effective theory. We leave the discussion on the proper incorporation of the *μe* correlation into the effective theory for a future study.

The paper is organized as the following. In section II the effective formulation of MP2 and CC theories are discussed. The computational implementation is detailed in Section III. In Section IV at the first part, the effective post-EHF methods are benchmarked on a diverse set of small muonic species whereas at the next step they are applied to a simple but representative set of muoniated N-Heterocyclic carbene/silylene/germylene species as well as the triazolium cation. Finally, our conclusions are offered in Section V.

## II. Theory

The two-component Coulombic Hamiltonian in atomic units for a system containing $N$ electrons, a single muon with mass $m_\mu$, and $q$ clamped nuclei with charges $Z_\beta$, all distinguished by their position vectors, is the following:

$$\hat{H}_{total} = \hat{H}_{NEO} + U_{classic}, \qquad U_{classic} = \sum_{\beta}^{q}\sum_{\gamma\rangle\beta}^{q} \frac{Z_\beta Z_\gamma}{\left|\vec{R}_\beta - \vec{R}_\gamma\right|}$$

$$\hat{H}_{NEO} = -(1/2)\sum_{i}^{N}\nabla_i^2 - (1/2\,m_\mu)\nabla_\mu^2 - \sum_{i}^{N}\frac{1}{\left|\vec{r}_\mu - \vec{r}_i\right|} + \sum_{i}^{N}\sum_{j\rangle i}^{N}\frac{1}{\left|\vec{r}_i - \vec{r}_j\right|} + V_{ext}$$

$$V_{ext} = -\sum_{i}^{N}\sum_{\beta}^{q}\frac{Z_\beta}{\left|\vec{R}_\beta - \vec{r}_i\right|} + \sum_{\beta}^{q}\frac{Z_\beta}{\left|\vec{R}_\beta - \vec{r}_\mu\right|} \qquad (1)$$

Neglecting the $\mu e$ correlation from the outset (and the spin variables for electrons just for brevity), the trial wavefunction ansatz for the ground state of the aforementioned system is: $\Psi_{trial} = \psi_e(\vec{r}_1,...,\vec{r}_N)\psi_\mu(\vec{r}_\mu)$. One may try to find the best trial function using the variational principle while the difference with the exact solution: $\hat{H}_{NEO}\Psi_{exact} = E_{exact}\Psi_{exact}$, is the rigorous definition of the $\mu e$ correlation energy as discussed in detail by Cassam-Chenaï and coworkers.[81,82] However, let us transform the problem into the language of an effective Hamiltonian assuming that the proper functional form of $\psi_\mu(\vec{r}_\mu)$ is known and only the set of its parameters, $\{\omega_k\}$, has to be determined: $\psi_\mu(\vec{r}_\mu) = f(\vec{r}_\mu, \{\omega_k\})$. By incorporating this functional form into the variational integral of total energy, one arrives at:

$$E = \int d\vec{r}_1 ... \int d\vec{r}_N \int d\vec{r}_\mu \Psi_{trial} \hat{H}_{NEO} \Psi_{trial} = \left[ \int d\vec{r}_1 ... \int d\vec{r}_N \psi_e \hat{H}^e_{eff}(\{\omega_k\}) \psi_e \right] + U_{eff}(\vec{R}_\beta, \{\omega_k\})$$

$$\hat{H}^e_{eff} = (-1/2) \sum_i^N \nabla_i^2 + \sum_i^N \sum_{j>i}^N \frac{1}{|\vec{r}_i - \vec{r}_j|} - \sum_i^N \sum_\beta^q \frac{Z_\beta}{|\vec{R}_\beta - \vec{r}_i|} + \sum_i^N V^{\mu-e}_{eff,i}(\vec{r}_i; \{\omega_k\})$$

$$U_{eff} = T(\{\omega_k\}) + \sum_\beta^q V^{\mu-nuc}_{eff,\beta}(\vec{R}_\beta, \{\omega_k\}) \qquad (2)$$

The three terms of: $T$, $V^{e-\mu}_{eff,i}$ and $V^{\mu-nuc}_{eff,\beta}$ originate from muon's kinetic energy integral, the muon-electron potential energy integral, and the muon-clamped nucleus potential energy integral, respectively.[56] One may re-interpret equation (2) as the variational integral for the ground state energy of the following effective Schrödinger equation: $\hat{H}^e_{eff}(\{\omega_k\}) \psi_{eff}(\{\omega_k\}) = E^e_{eff}(\{\omega_k\}) \psi_{eff}(\{\omega_k\})$. Formally, the optimization of the total energy: $E_{total} = E^e_{eff} + U_{eff}$, with respect to $\{\omega_k\}$ using the exact solution of the effective Schrödinger equation is equal to finding the best energy-optimized $\Psi_{trial}$ as proposed above. In this viewpoint the set of $\{\omega_k\}$ is determined via the optimization procedure that is employed for the optimization of the nuclear geometry. At the first stage, the ground state wavefunction of a 3D-isotropic harmonic oscillator, $f(\vec{r}_\mu, \{\alpha, \vec{R}_c\}) = (2\alpha/\pi)^{\frac{3}{4}} \exp(-\alpha |\vec{r}_\mu - \vec{R}_c|^2)$,[56] is adopted for the approximation of muon's distribution that leads to the following effective Hamiltonian:

$$\hat{H}^e_{eff} = (-1/2) \sum_i^N \nabla_i^2 + \sum_i^N \sum_{j>i}^N \frac{1}{|\vec{r}_i - \vec{r}_j|} - \sum_i^N \sum_\beta^q \frac{Z_\beta}{|\vec{R}_\beta - \vec{r}_i|} - \sum_i^N \frac{erf\left[\sqrt{2\alpha}|\vec{r}_i - \vec{R}_c|\right]}{|\vec{r}_i - \vec{R}_c|}$$

$$U_{eff}\left(\vec{R}_\beta,\{\alpha,\vec{R}_c\}\right)=(3\alpha/2m_\mu)+\sum_\beta^q \frac{Z_\beta\, erf\left[\sqrt{2\alpha}\left|\vec{R}_\beta-\vec{R}_c\right|\right]}{\left|\vec{R}_\beta-\vec{R}_c\right|} \quad (3)$$

More complicated effective Hamiltonians are derivable if more complex oscillator models are employed for muon's vibrations.[56]

In order to proceed further let us assume that $\psi_e$, for a typical closed-shell system, is a Slater determinant formed from $N$ electronic spin-orbitals constructed from $N/2$ spatial orbitals, $\{\psi_j\}, j=1,...,N/2$. The conventional functional variation of equation (2): $\delta_{\{\psi_j\}}E=0$,[43] after some mathematical manipulations, yields the following set of the HF differential equations:

$$\hat{f}_{eff}\psi_j(\vec{r}_1)=\varepsilon_j\psi_j(\vec{r}_1)$$

$$\hat{f}_{eff}=(-1/2)\nabla_1^2-\sum_\beta^q \frac{Z_\beta}{\left|\vec{R}_\beta-\vec{r}_1\right|}+\sum_j^{N/2}\left(2\hat{J}_j(\vec{r}_1)-\hat{K}_j(\vec{r}_1)\right)+V_{eff,1}^{\mu-e}(\vec{r}_1;\{\omega_k\}) \quad (4)$$

These are the EHF equations that were also derived in the previous studies while $\hat{J}$ and $\hat{K}$ are the conventional Coulomb and exchange operators, respetcively.[55,56] Employing just a single Slater determinant is the simplest approximation for $\psi_e$ in the conventional ab initio electronic structure theory. More complicated approximations, i.e. a linear combination of Slater determinants, may be employed in equation (2) to derive the effective configuration interaction (ECI) or the effective multi-configurational self-consistent field (EMCSCF) theories.[43,44] However in this study we concentrate on non-variational improvements of the EHF results through employing the CC and the MP2 theories.

Herein, we do not discuss the conventional single-component formulation of the CC and the MP2 theories within the context of the adiabatic electronic structure theory because they are now part of the classics of quantum chemistry.[43,44,83-89] Both the MP2,[90-93] and the CC,[94-97] theories have been extended also for the multi-component systems and particularly, the multi-component MP2 method has been applied in many recent studies.[98-109] In the multi-component MP2 formalism where the NEO-MP2 is a typical exmaple,[90] the two separate correction terms for the muonic systems include the second order *ee* and *μe* correlation energies and is called NEO-MP2(*ee*+*μe*). Although the formalism is theoretically sound, in computational studies the *μe* correction term does not always yield reliable results (Goli and Shahbazian, unpublished results), casting some doubt on whether the NEO-MP2 is always a proper tool for recovering the *μe* correlation energy. Theoretically, higher order MP perturbative corrections may be invoked to remedy this problem, however, some classic studies demonstrate that the electronic MP perturbation series usually oscillates and even diverge in non-trivial manners.[110,111] Since the electron-electron interaction term is unaltered in the effective Hamiltonian, and the EHF theory contains the same Coulomb and exchange operators as those of the conventional HF theory, the effective MP2 (EMP2) correction term for a closed-shell system is formally equal to that of the NEO-MP2(*ee*):

$$E_{EMP2} = E_{EHF} + \sum_{a,b=1}^{N/2} \sum_{r,s=(N/2+1)}^{K} \frac{\langle ab|rs\rangle \left(2\langle rs|ab\rangle - \langle rs|ba\rangle\right)}{\varepsilon_a + \varepsilon_b - \varepsilon_r - \varepsilon_s}$$

$$\langle ij|kl\rangle = \int d\vec{r}_1 \int d\vec{r}_2 \, \psi_i^*(\vec{r}_1)\psi_j^*(\vec{r}_2)\left(1/|\vec{r}_1 - \vec{r}_2|\right)\psi_k(\vec{r}_1)\psi_l(\vec{r}_2) \qquad (5)$$

In this equation $a,b$ and $r,s$ stand for the occupied and the virtual EHF orbitals, respectively, while $K$ is the number of virtual orbitals originating from the algebraic solution of the EHF equations. The same reasoning is also applicable to the CC theory as well and the correction terms are formally equal to those of the conventional single-component theory.[84-89] The only difference, as is also evident from equation (5), is the replacement of the occupied and virtual orbitals of the conventional HF method with those derived from the EHF equations. In the present study both standard CC truncations namely the CCSD,[112,113] and the CCSD(T),[114] are also employed and termed as the effective CCSD (ECCSD) and the effective CCSD(T) (ECCSD(T)) methods. Let us finally stress that the recent multi-component CC studies of Udagawa and Tachikawa is similar in its spirit to our proposed effective CC formulation though the authors did not explicitly offer their formalism.[115,116]

## III. Computational implementation

The details of the EHF calculations and computational procedure have been disclosed comprehensively in the previous study and are not reiterated herein.[56] The same electronic closed-shell set of hydrides used in the previous study was also employed for a diverse benchmarking in the present study which includes: LiH, $BeH_2$, $BH_3$, $CH_4$, $NH_3$, $H_2O$, FH, NaH, $MgH_2$, $AlH_3$, $SiH_4$, $PH_3$, $H_2S$, HCl. Similar to the previous study,[56] in each of these species one of the hydrogen nuclei was replaced by a muon as a quantum particle (with the mass of 206.768 in atomic units), where it was represented by a single *ghost* atom. The rest of nuclei were treated as clamped point charges (instead of H, µ is used to represent this ghost atom, e.g. LiMu and BeHMu). In all calculations the correlation consistent aug-cc-pVTZ electronic basis set (with Cartesian basis functions) was used for electrons at the clamped and the ghost centers; in principle the correlation consistent hierarchy may be specifically re-designed for the muon, however, in the present study aug-cc-pVTZ basis set of hydrogen atom was employed instead.[117,118] The effective potential was derived from the uncontracted muonic [2s2p2d] basis set where the exponents were fixed at their recommended values offered in the previous study (See Table 1 in Ref. [56] for the numerical values of all the exponents). Thus, only the linear coefficients in the effective potential were optimized during the SCF cycles of the EHF calculations (for a comprehensive discussion on the details see the supporting information (SI) of Ref. [56]]). All molecular geometries, i.e. the positions of the clamped nuclei and the ghost atom, were optimized using a numerical approximation of the post-EHF energy gradients. Because the muon is considered as a light quantum particle with anharmonic vibrations, as discussed in the previous study,[56] the direct distance between the ghost atom and the muon binding center is not a realistic measure of the muonic bond length. Thus, the mean inter-nuclear distance was computed as the expectation value of the position operator with respect to the muon addition center, using the optimized muonic orbital as the wavefunction. For comparison purposes, the set of hydrogen congeners were also optimized within the adiabatic paradigm, i.e. assuming all nuclei as clamped point charges, at the conventional HF, MP2, CCSD and CCSD(T) levels of theory. In the case of muoniated adducts of N-Heterocyclic carbene/silylene/germylene species and triazolium cation the restricted open-shell EKS equations with B3LYP electronic functional,[119-121] with aug-cc-pVTZ basis set were employed for geometry optimization (termed RO-EB3LYP/aug-cc-pVTZ). The details of the computational implementation of the EKS equations are discussed in detail previously.[57] At the next stage single-point calculations were done on the optimized geometries using the restricted open-shell version of the post-EHF methods, namely, RO-EMP2,[122,123] and RO-ECCSD,[124,125] by a modified version of the GAMESS package.[126,127]

## IV. Results and discussion

### A. Benchmark calculations

Let us start by considering the muonic bond lengths computed at the post-EHF levels that are among distinctive features of the muonic species since they are significantly different from those of the analog hydrides.[56] Table 1 offers the muonic bond lengths whereas Table S1 in the SI provides the same results for hydrides. Even a brief glance at Tables 1 and S1 reveals that in contrast to some hydrides, the mean muonic bond lengths are not sensitive to the *ee* correlation and the results at the EHF level are marginally altered at the considered post-EHF levels. Accordingly, Figure 1 depicts the differences of the muonic and analog bond lengths in hydrides from Tables 1 and S1 (for numerical values see Table S2 in the SI). The most notable feature observable in this comparative analysis is the non-negligible elongation of the muonic bond lengths versus the analog bond lengths in the hydrides at all computational levels, which is a general characteristics of the muonic species.[56] This difference is larger for species with central atoms from the left-hand side of the Periodic Table (PT) and diminishes for species with electronegative central atoms from the right-hand side of the PT. In the case of the hydrides containing central atoms from the last three groups of the table, i.e. $5^{th}$ to $7^{th}$ columns, the role of the *ee* correlation is not marginal as for (FH,FMu) and (HCl,MuCl) pairs the differences computed at CCSD(T)/ECCSD(T) levels are ~0.02 Å less than those computed at HF/EHF levels. Evidently, the difference is exaggerated at the mean-field level and this cast some doubt on using a single universal scaling factor to estimate the muonic bond lengths from their hydrogenic analog regardless of the used computational level. Additionally, Table S3 in the SI offers the bond lengths of the vicinal bonds between the clamped hydrogen and the central atoms in the muonic species. They are virtually indistinguishable from the bond lengths computed for the congener hydrides and one may conclude that the substitution of a clamped proton with a muon imperceptibly affects the vicinal bond lengths. In order to check the variations of the *ee* correlation energy upon subsituating a clamped proton with a muon in a moleuclar species, Table 2 discloses the computed correlation energies for both muonic speices and the hydrides (for total energies see Table S4 in the SI). Also, Figure 2 depcits the variations at each computational level across each row of the PT. Even a fast glance on the table and the figure reveals that not only the general pattern of variations within the rows but also the absolute amounts of the *ee* correlation energies are quite similar when each pair of the muonic and corresponding hydride species is compared. Clearly, the effective potential has a marginal role in the values of the *ee* correlation energy and is not senstive to the substitution of the clamped particle with a light quantum particle.

Table 1- The mean muonic bond lengths computed at both EHF and the post-EHF levels (in Angstroms).

|            | EHF   | EMP2  | ECCSD | ECCSD(T) |
|------------|-------|-------|-------|----------|
| **LiMu**   | 1.695 | 1.690 | 1.692 | 1.692    |
| **BeHMu**  | 1.413 | 1.410 | 1.411 | 1.411    |
| **BH$_2$Mu** | 1.264 | 1.257 | 1.262 | 1.262    |
| **CH$_3$Mu** | 1.151 | 1.148 | 1.149 | 1.150    |
| **NH$_2$Mu** | 1.059 | 1.058 | 1.058 | 1.058    |
| **OHMu**   | 0.997 | 0.997 | 0.997 | 0.997    |
| **FMu**    | 0.953 | 0.954 | 0.954 | 0.954    |
| **NaMu**   | 1.998 | 1.993 | 1.998 | 1.998    |
| **MgHMu**  | 1.790 | 1.784 | 1.787 | 1.787    |
| **AlH$_2$Mu** | 1.664 | 1.657 | 1.659 | 1.660    |
| **SiH$_3$Mu** | 1.559 | 1.553 | 1.555 | 1.555    |
| **PH$_2$Mu** | 1.488 | 1.484 | 1.485 | 1.486    |
| **SHMu**   | 1.406 | 1.400 | 1.401 | 1.401    |
| **MuCl**   | 1.341 | 1.336 | 1.336 | 1.336    |

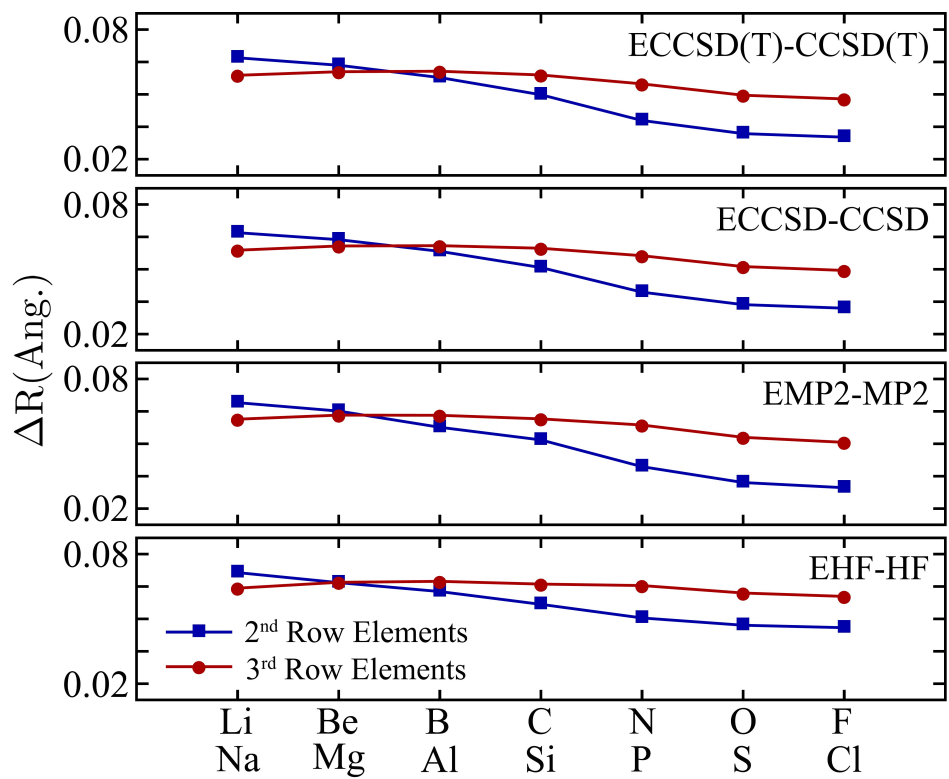

Figure 1- The difference (in Angstroms) between the mean muonic bond lengths (see Table 1) and the analog bond lengths in the hydrides (see Table S1).

Table 2- The correlation energies at various post-EHF and post-HF levels (in Hartrees) for the muonic speices and the hydrides, respectively.

|  | EMP2 | ECCSD | ECCSD(T) |  | MP2 | CCSD | CCSD(T) |
|---|---|---|---|---|---|---|---|
| **LiMu** | -0.0269 | -0.0348 | -0.0348 | **LiH** | -0.0284 | -0.0365 | -0.0365 |
| **BeHMu** | -0.0613 | -0.0765 | -0.0772 | **BeH$_2$** | -0.0620 | -0.0771 | -0.0771 |
| **BH$_2$Mu** | -0.1156 | -0.1370 | -0.1398 | **BH$_3$** | -0.1153 | -0.1367 | -0.1393 |
| **CH$_3$Mu** | -0.2034 | -0.2225 | -0.2294 | **CH$_4$** | -0.2019 | -0.2217 | -0.2217 |
| **NH$_2$Mu** | -0.2439 | -0.2543 | -0.2632 | **NH$_3$** | -0.2413 | -0.2528 | -0.2614 |
| **OHMu** | -0.2732 | -0.2764 | -0.2856 | **H$_2$O** | -0.2704 | -0.2747 | -0.2837 |
| **FMu** | -0.2854 | -0.2854 | -0.2933 | **FH** | -0.2829 | -0.2839 | -0.2917 |
| **NaMu** | -0.0268 | -0.0353 | -0.0353 | **NaH** | -0.0285 | -0.0374 | -0.0374 |
| **MgHMu** | -0.0577 | -0.0735 | -0.0740 | **MgH$_2$** | -0.0589 | -0.0748 | -0.0753 |
| **AlH$_2$Mu** | -0.0961 | -0.1191 | -0.1209 | **AlH$_3$** | -0.0965 | -0.1196 | -0.1213 |
| **SiH$_3$Mu** | -0.1472 | -0.1764 | -0.1807 | **SiH$_4$** | -0.1469 | -0.1762 | -0.1803 |
| **PH$_2$Mu** | -0.1754 | -0.2025 | -0.2101 | **PH$_3$** | -0.1743 | -0.2021 | -0.2022 |
| **SHMu** | -0.1971 | -0.2194 | -0.2283 | **H$_2$S** | -0.1959 | -0.2193 | -0.2280 |
| **MuCl** | -0.2099 | -0.2283 | -0.2371 | **HCl** | -0.2091 | -0.2286 | -0.2373 |

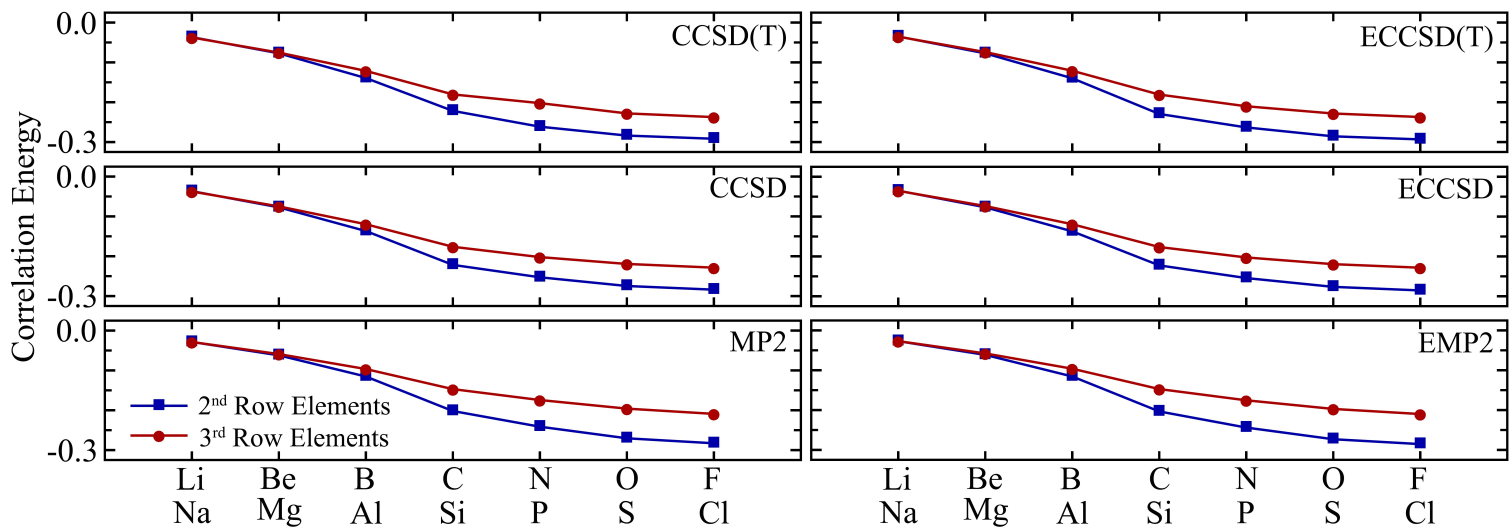

Figure 2- The correlation energies (in Hartrees) of the muonic species and the hydrides at various post-EHF and post-HF levels, respectively.

### B. Applications

The results gained from the benchmark set though hard to be rationalized based on a simplified model, points to the fact that accurate computation of the *ee* correlation at the post-EHF levels is not crucial to determine the muonic bond length. Thus, in this section we apply the previously proposed EKS equations with the B3LYP electronic exchange-correlation functional to optimize the geometries of the muonic species and then the post-EHF single-point calculations to evaluate the relative stabilities of various muonic systems.

In order to examine the real-life applicability of the developed effective methodologies, the parent N-Heterocyclic carbene (NHC) and its heavier analogs, i.e. the parent N-Heterocyclic silylene (NHSi) and germylene (NHGe) species as well as the triazolium cation (TAC) are considered in this section. The general structure of the studied molecules is depicted in Scheme 1.

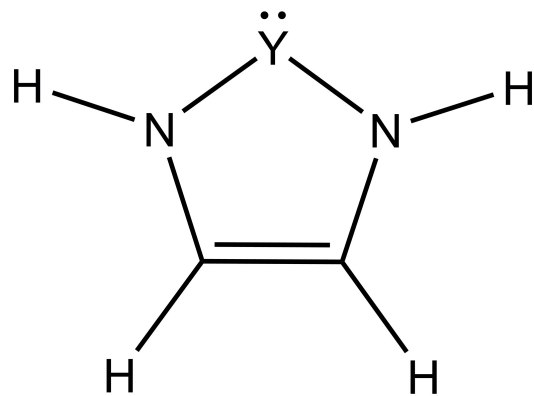

Scheme 1- The schematic molecular structure of the considered species (Y = C, N$^+$, Si, Ge)

These species serve as models for the NH divalent species, with usually large substituents attached to the nitrogen atoms, that have been considered by the muon spin resonance spectroscopy techniques in recent years.[19,20,128-136] Most of these divalent species were once considered as unstable and inaccessible but were eventually synthesized and found vast applications in various fields of chemistry. Some classic reviews on the original studies are Ref. 137-139 while more recent developments may be found in Ref. 140-142, for a brief but vivid historical overview see Ref. 143, and for having a broader perspective on the field of carbenes and their heavier analogs Ref. 144-147 are recommendable. To the best of authors' knowledge, the triazolium cation has not been considered by the muon spin resonance spectroscopy but it is a well-known member of the nitrenium cations.[148-150] Let us just stress that although the parent species are not known as bulk stable compounds, at least in the case of the parent NHC it is detectable in molecular beams and matrix-isolated low-temperature experiments and also through computational studies.[151-153] Not only the corresponding muoniated species are interesting in themselves, but also their examination may offer insights into the reaction centers and the details of the radical reactions in the NH divalent species.[154]

Figure 3 depicts the optimized structures of the parent molecules at B3LYP/aug-cc-pVTZ level with some geometrical parameters included; the Cartesian coordinates are offered in the SI. The computed geometrical parameters are completely in line with the previous report,[153] reproducing the well-known trends among the NH species.[137-141] At the next step, a hydrogen atom was added to each parent structure in order to model the attachment of the muonium atom to these molecules, as is common in the relevant literature.[129-136] In principle, the hydrogen atom may be attached to the divalent center or to one of the "unsaturated" carbons with a double bond in between resulting in two distinct adducts.

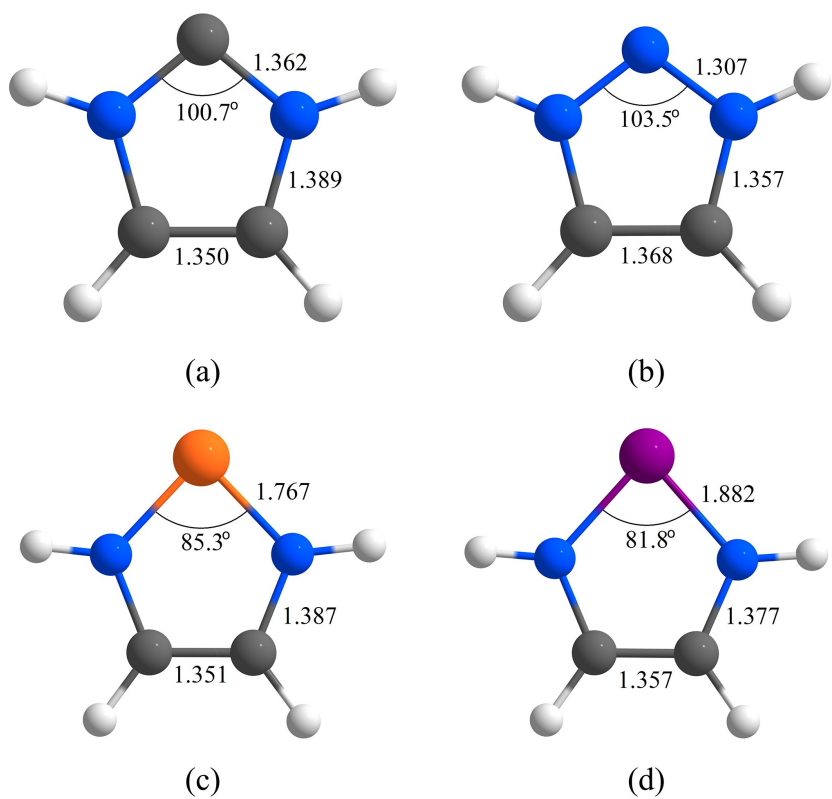

Figure 3- The optimized geometries (distances in Angstroms) of the four considered planar parent divalent species, a) NHC, b) TAC, c) NHSi, d) NHGe, derived at B3LYP/aug-cc-pVTZ level. The carbon, silicon, germanium, nitrogen and hydrogen atoms are depicted as gray, orange, purple, blue and white spheres.

Each adduct is named by the following convention: the attached hydrogen atom to the parent molecule is given first followed by the name of the atom to which it is attached, and finally, the parent molecule name itself is indicated, e.g. H-Si-NHSi means an adduct of hydrogen atom addition to Si center of the NHSi. In the case of the NHC, to prevent ambiguity, the carbeneic carbon is denoted by cC while alkeneic carbon is designated as aC, e.g. H-aC-NHC means an adduct of hydrogen attachment to one of the two alkeneic carbons of the NHC. Subsequently, the same naming conventions are also used for the muonic adducts, where only H is replaced by Mu, e.g. Mu-aC-NHC. Figure 4 represents the optimized structures of the hydrogen adducts derived at RO-B3LYP/aug-cc-pVTZ level while the Cartesian coordinates may be found in the SI. The trends observed in this figure are similar to previous studies and are discussed herein only briefly.[129,155,156] When the hydrogen atom is added to the divalent center, this mostly influences the lengths of the adjacent divalent center-nitrogen bonds yielding a considerable elongation. Concomitantly, the angle between the molecular plane, containing nitrogen atoms and the divalent center, and the bond between th divalent center and hydrogen atom increases in the case of the larger divalent atoms.[128] On the other hand, when the hydrogen atom is added to the alkeneic atom, the adjacent C-C and C-N bonds significantly elongate while the corresponding N-C-C bond angle shrinks, which are both in line with a formal change of the carbon's hybridization from $sp^2$ to $sp^3$. At this stage we may compare the results gained from the addition of the muonium atom to the parent species at RO-EB3LYP/aug-cc-pVTZ level with those of the hydrogen adducts. Figure 5 illustrates the optimized structures of the muoniated adducts and some geometrical parameters while the Cartesian coordinates may be found in the SI. The most prominent pattern of alternation is the substantial elongation, 0.07 - 0.10 Å, in the mean muonic bond lengths relative to their hydrogenic counterparts, which are in line with the previous studies.[56,57] Interestingly, none of the remaining geometrical parameters show large variations relative to their hydrogenic congeners, demonstrating that apart from the muonic bond length, hydrogen atom may serve as a good model for the muonium atom addition to the studied species. It is also interesting to check the electronic structure differences among the muoniated adducts that may be used in future studies to interpret the muon spin resonance spectrum (we postpone a quantitative analysis after the introduction of the $\mu e$ correlation into the effective theory and direct computation of the hyperfine coupling constants). Figure 6 offers the singly occupied molecular orbital (SOMO) of the muoniated species at RO-EB3LYP/aug-cc-pVTZ level (the SOMOs of the hydrogen adducts are practically indistinguishable from the muonic counterparts and are not depicted). It is clear from panels of Figure 6(a-d) that the SOMOs of these species are quite distinct from each other and while in the case of Mu-cC-NHC and Mu-N-TAC the unpaired electron mostly prefers the divalent center and its neighboring nitrogen atoms, for Mu-Ge-NHGe it resides at the alkeneic carbon atoms instead of the divalent center. In panels of Figure 6(e-h) less dramatic differences are observable and the excessive electron is mostly located at the carbon atom adjacent to the addition site, although Mu-C-TAC shows a distinct and more delocalized pattern of the spin density. In order to study the relative stabilities of the muoniated species Tables 3 and S5 in the SI list the relative and total energies, respectively. The ground state energies are derived at RO-EB3LYP/aug-cc-pVTZ level.

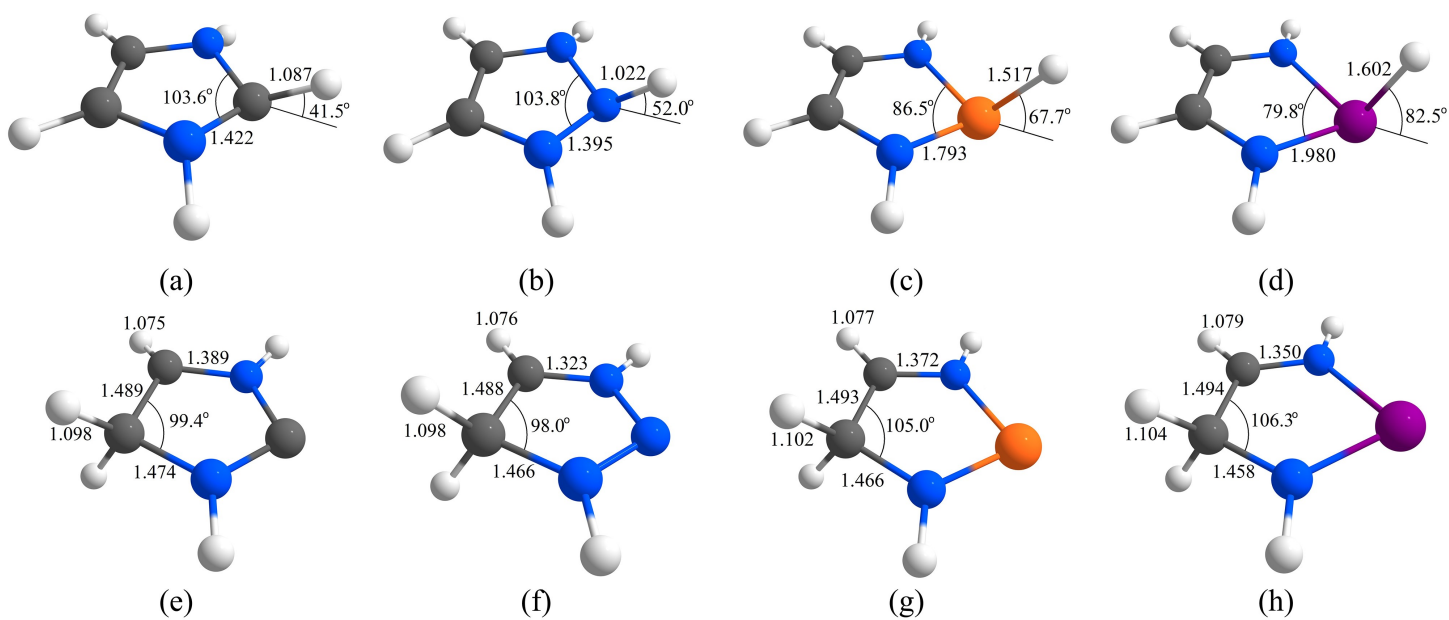

Figure 4- The optimized geometries (distances in Angstroms) of the eight adducts resulting from adding hydrogen atom to the parent divalent species, a) H-cC-NHC, b) H-N-TAC, c) H-Si-NHSi, d) H-Ge-NHGe, H-aC-NHC, f) H-C-TAC, g) H-C-NHSi, h) H-C-NHGe, all computed at RO-B3LYP/aug-cc-pVTZ level. The carbon, silicon, germanium, nitrogen and hydrogen atoms are depicted as gray, orange, purple, blue and white spheres.

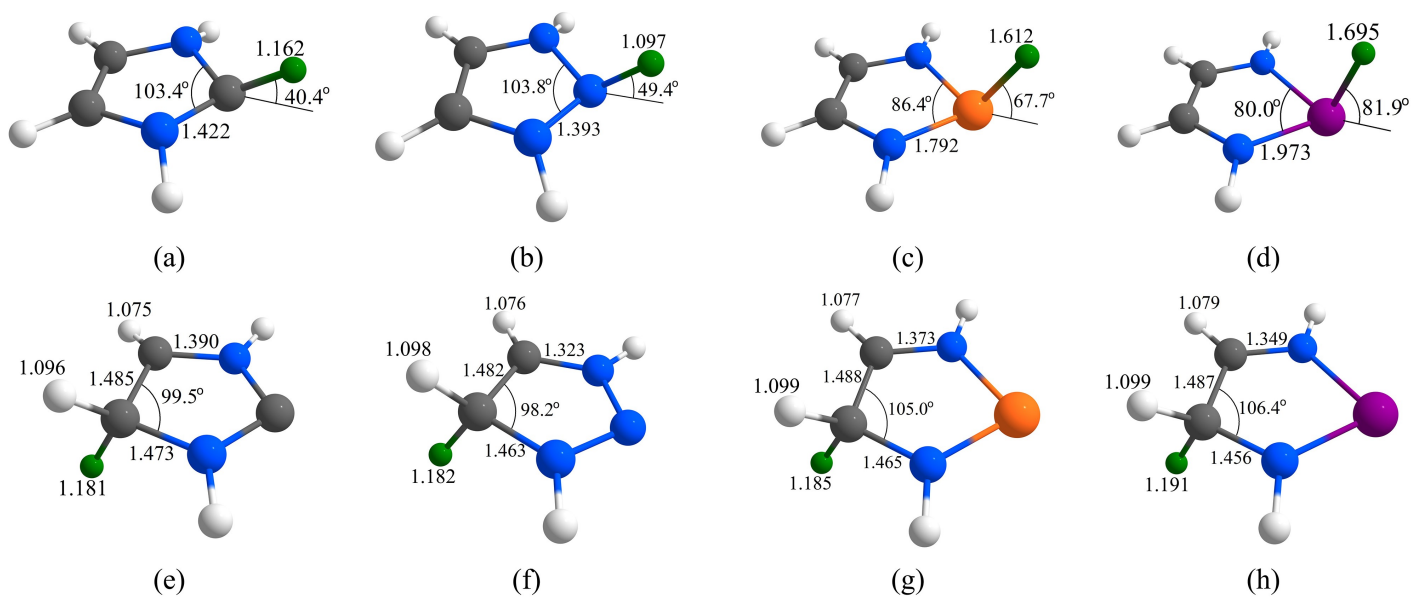

Figure 5- The optimized geometries (distances in Angstroms) of the eight adducts resulting from adding muonium atom to the parent divalent species, a) Mu-cC-NHC, b) Mu-N-TAC, c) Mu-Si-NHSi, d) Mu-Ge-NHGe, Mu-aC-NHC, f) Mu-C-TAC, g) Mu-C-NHSi, h) Mu-C-NHGe, all computed at RO-EB3LYP/aug-cc-pVTZ level. The carbon, silicon, germanium, nitrogen, hydrogen and muonium atoms are depicted as gray, orange, purple, blue, white and green spheres.

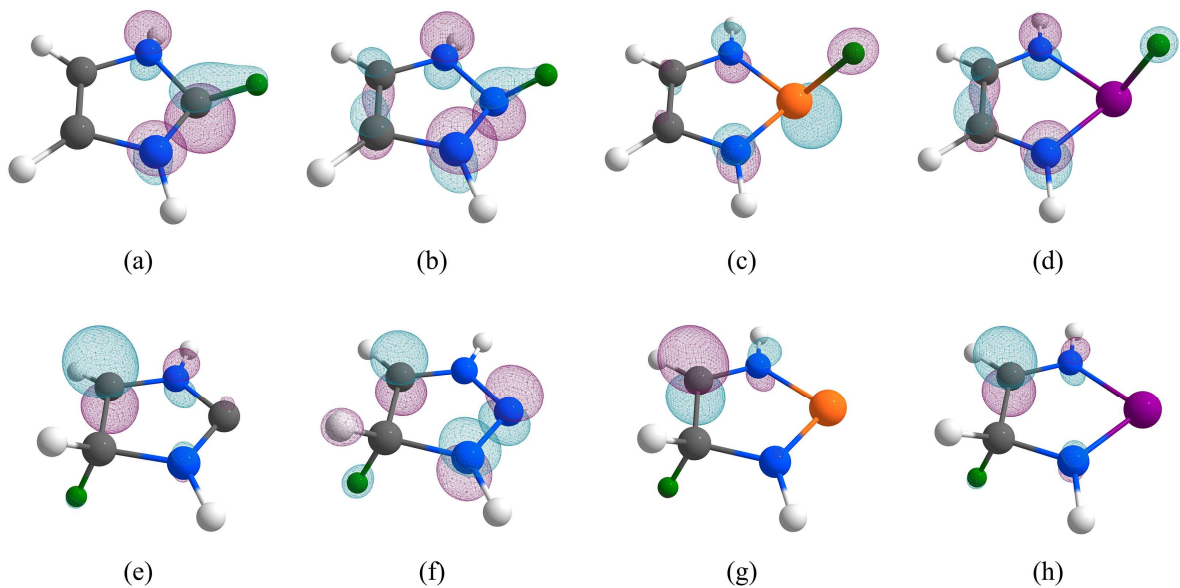

Figure 6- The SOMOs of the eight adducts resulting from the addition of the muonium atom to the parent divalent species, a) Mu-cC-NHC, b) Mu-N-TAC, c) Mu-Si-NHSi, d) Mu-Ge-NHGe, Mu-aC-NHC, f) Mu-C-TAC, g) Mu-C-NHSi, h) Mu-C-NHGe, at RO-EB3LYP/aug-cc-pVTZ level (the used iso-surface is 0.1 in atomic units). The carbon, silicon, germanium, nitrogen, hydrogen and muonium atoms are depicted as gray, orange, purple, blue, white and green spheres.

Single-point calculations were performed at RO-EMP2/aug-cc-pVTZ and RO-ECCSD/aug-cc-pVTZ levels employing the optimized geometries at RO-EB3LYP/aug-cc-pVTZ level, i. e. RO-EMP2/aug-cc-pVTZ//RO-EB3LYP/aug-cc-pVTZ and RO-ECCSD/aug-cc-pVTZ//RO-EB3LYP/aug-cc-pVTZ; for brevity, these levels are termed as EB3LYP, EMP2 and ECCSD in Table 3. In order to perform a contrastive study, the hydrogen adduct energetic data are also provided in the same tables at comparable levels of theory, i.e. RO-B3LYP/aug-cc-pVTZ, RO-MP2/aug-cc-pVTZ//RO-B3LYP/aug-cc-pVTZ, CCSD/aug-cc-pVTZ//RO-B3LYP/aug-cc-pVTZ; for brevity, these levels are indicated as B3LYP, MP2, CCSD in Table 3, respectively. Evidently, irrespective to the used computational level, in all cases the addition of hydrogen or the muonium atom to the divalent center is more energetically favored than the addition to alkeneic center, which is completely in line with previously derived computational and experimental results.[128-136]

Table 3- The relative energies of each pair of the muoniated (and the hydrogenated) adducts (in kJ/mol).*

|  | B3LYP | MP2 | CCSD |  | EB3LYP | EMP2 | ECCSD |
|---|---|---|---|---|---|---|---|
| **H-aC-NHC** | 58.2 | 70.5 | 50.3 | **Mu-aC-NHC** | 58.5 | 71.4 | 50.7 |
| **H-C-TAC** | 15.5 | 39.4 | 8.8 | **Mu-C-TAC** | 22.3 | 51.7 | 16.0 |
| **H-C-NHSi** | 23.5 | 21.4 | 4.2 | **Mu-C-NHSi** | 41.5 | 38.3 | 22.3 |
| **H-C-NHGe** | 16.7 | 13.8 | 3.0 | **Mu-C-NHGe** | 39.7 | 36.9 | 26.4 |

* In each group of the same parent species, the most stable system is used as the reference (zero) point of the energy and the less stable system is given as the entry.

However, apart from the NHC adducts where the relative energies of the hydrogen and the muoniated adducts are similar, in the case of the three remaining parents the muonium atom more strongly favors the divalent center. This disparity between the hydrogen and the muoniated adducts at first glance seems to be odd since the geometrical differences, depicted in Figures 4 and 5, are mainly confined to the lengths of the muonic and hydrogenic bonds to the addition centers ($R_{Mu-X}-R_{H-X} = 0.85 \pm 0.15$ Å). However, this distinction does not seem to depend on the geometrical differences, and instead, has to be traced in subtle distortions of the electronic density upon substitution of the proton with the muon. Table 4 presents the atomic charges of the added hydrogen and the muonium atoms, as one of the simplest local measures of the electronic distribution, using the quantum theory of atoms in molecules (QTAIM) methodlogy,[157] and its multi-component version (MC-QTAIM),[158] computed from the one-particle densities derived at RO-B3LYP/aug-cc-pVTZ and RO-EB3LYP/aug-cc-pVTZ levels, respectively. It is noteworthy that the MC-QTAIM analysis reveals the regional traits of the electronic and the muonic real-space distributions in the muonic species, in addition to their AIM structure.

Table 4- The atomic charges, q, of the added hydrogen and the muonium atoms for all the hydrogenated and the muoniated adducts. The difference of the atomic charges, Δq, between the hydrogen atoms (and the muonium atoms) of each pair of adducts is also given.

|  | H-cC-NHC | H-N-TAC | H-Si-NHSi | H-Ge-NHGe |
|---|---|---|---|---|
| q | 0.02 | 0.46 | -0.64 | -0.40 |
|  | H-aC-NHC | H-C-TAC | H-C-NHSi | H-C-NHGe |
| q | 0.01 | 0.12 | -0.01 | -0.01 |
| Δq | 0.01 | 0.34 | -0.63 | -0.39 |
|  | Mu-cC-NHC | Mu-N-TAC | Mu-Si-NHSi | Mu-Ge-NHGe |
| q | 0.28 | 0.66 | -0.56 | -0.31 |
|  | Mu-aC-NHC | Mu-C-TAC | Mu-C-NHSi | Mu-C-NHGe |
| q | 0.24 | 0.37 | 0.21 | 0.20 |
| Δq | 0.04 | 0.28 | -0.77 | -0.51 |

The MC-QTAIM analysis has been previously applied to various muonic molecules, and unraveled the distinct features of the bonded muonium atoms.[50-53] The most important feature observable from the data in Table 4, in line with the previous studies,[50-53] is the lower intrinsic ability of the muonium atoms to retain electrons in their own atomic basins relative to their counterpart hydrogen atoms. The more interesting pattern, however, emerges if one compares the differences in atomic charges between each pair of adducts. In the case of the NHC adducts, the difference of atomic charges of the added hydrogen atoms in H-cC-NHC and H-aC-NHC pair, ~0.01e, is quite similar to that of the muonium atoms in Mu-cC-NHC and Mu-aC-NHC pair, ~0.04e. This observation reveals the fact that the local electronic environment of the added hydrogen and the muonium atoms is not significantly different in each pair of adducts; this is in part understandable since the added hydrogen and the muonium atoms in each pair encounter only carbon atoms as their immediate neighbors. However, this is not the case for remaining adducts and the differences in atomic charges are large and distinct if one compares the hydrogen and the muoniated adducts for each group of the parent species. This trend correlates well with the stability trends observed in Table 3 revealing a delicate difference between hydrogen and the muonium atoms when bonded in a chemical environment. Thus, modelling muonium addition by using a clamped hydrogen atom may not always reproduce the correct trends in relative stabilities and does not seem to be a generally safe approach. Evidently, in contrast to their similarities, hydrogen and muonium atoms have to be conceived as "chemically" distinct species with delicate, but, distinguishable characteristics.

## V. Conclusion

The present study is part of our ongoing programme to devise a muonic-specific ab initio electronic theory that tackles the computational study of the muonic molecules, trying to combine theoretical simplicity and computational feasibility.[55-57] While it is customary to use the conventional adiabatic electronic structure theory to model the muonic molecules,[21,129-136,159-161] assuming that a hydrogen atom with a clamped nucleus may be used to model the muonium atom, our studies do not confirm that this viewpoint is always a legitimate computational strategy. On the contrary, the present study demonstrates that both muonic bond lengths and the relative stabilities of the muonic adducts are non-negligibly underestimated within context of the adiabatic strategy. While, in principle, it is possible to employ certain *ad hoc* corrections to remedy the shortcomings of the adiabatic strategy in a specific group of the muonic molecules, it seems hard to contemplate how it is possible to devise a universal and theoretically well-founded adiabatic strategy to model the muonic systems. On the other hand, our own results also suffer from the lack of proper treatment of the *μe* correlation and this need to be taken into account if one pursues the "holy grail" of computational muonic chemistry, namely, quantitative prediction of the hyperfine coupling constants and the muon spin resonance spectrum. Thus, before further computational developments it is desirable to include, even at a crude level, the *μe* correlation into the effective theory. This line of research is currently under study in our lab and the corresponding results will be reported in future.

## Acknowledgments

The authors are grateful to Cina Foroutan-Nejad for detailed reading of this paper.

# References

1. D. C. Walker, *Muon and Muonic Chemistry* (Cambridge University Press, Cambridge, 1983).

2. E. Roduner, *The Positive Muon as a Probe in Free Radical Chemistry* (Lecture Notes in Chemistry, Vol. 49, Springer, Berlin, 1988).

3. S. J. Blundell, Contemp. Phys. **40**, 175 (1999).

4. S. J. Blundell, Chem. Rev. **104**, 5717 (2004).

5. N. J. Clayden, Phys. Scr. **88**, 068507 (2013).

6. L. Nuccio, L. Schulz and A. J. Drew, J. Phys. D: Appl. Phys. **47**, 473001 (2014).

7. J. Phys. Soc. Jap. **85**, No. 9 (2016), Special Topics: Recent Progress of Muon Science in Physics, 091001-091016.

8. E. Roduner, Chem. Soc. Rev. **22**, 337 (1993).

9. E. Roduner, Appl. Magn. Reson. **13**, 1 (1997).

10. D. C. Walker, J. Chem. Soc., Faraday Trans. **94**, 1 (1998).

11. C. J. Rhodes, J. Chem. Soc. Perkin Trans. 2 1379 (2002).

12. I. McKenzie and E. Roduner, Naturwissenschaften **96**, 873 (2009).

13. I. McKenzie, Annu. Rep. Prog. Chem., Sect. C: Phys. Chem. **109**, 65 (2013).

14. L. Nuccio, L. Schulz, and A. J. Drew, J. Phys. D: Appl. Phys. **47**, 473001 (2014).

15. K. Ghandi and A. MacLean, Hyperfine Int. **230**, 17 (2015).

16. J. H. Brewer, Phys. Proc. **30**, 2 (2012).

17. K. Nagamine, *Introductory Muon Science* (Cambridge University Press, 2003).

18. U. A. Jayasooriya, R. Grinter, P. L. Hubbard, G. M. Aston, J. A. Stride, G. A. Hopkins, L. Camus, I. D. Reid, S. P. Cottrell and S. F. J. Cox, Chem. Eur. J. **13**, 2266 (2007).

19. A. Mitra, J.-C. Brodovitch, C. Krempner, P. W. Percival, P. Vyas and R. West, Angew. Chem. Int. Ed. **49**, 2893 (2010).

20. R. West, K. Samedov and P. W. Percival, Chem. Eur. J. **20**, 9184 (2014).

21. I. McKenzie, Phys. Chem. Chem. Phys. **16**, 10600 (2014).

22. J.-C. Brodovitch, B. Addison-Jones, K. Ghandi, I. McKenzie and P. W. Percival, Phys. Chem. Chem. Phys. **17**, 1755 (2015).

23. K. Wang, P. Murahari, K. Yokoyama, J. S. Lord, F. L. Pratt, J. He, L. Schulz, M. Willis, J. E. Anthony, N. A. Morley, L. Nuccio, A. Misquitta, D. J. Dunstan, K. Shimomura, I. Watanabe, S. Zhang, P. Heathcote and A. J. Drew, Nature Mat. **16**, 467 (2017).

24. C. G. Van de Walle, Phys. Rev. Lett. **64**, 669 (1990).

25. R. M. Valladares, A. J. Fisher, and W. Hayes, Chem. Phys. Lett. **242**, 1 (1995).

26. M. I. J. Probert, and A. J. Fisher, J. Phys.: Condens. Matter **9**, 3241 (1997).

27. A. R. Porter, M. D. Towler, and R. J. Needs, Phys. Rev. B **60**, 13534 (1999).

28. D. Cammarere, R. H. Scheicher, N. Sahoo, T. P. Das, and K. Nagamine, Physica B **289-290**, 636 (2000).

29. C. P. Herrero and R. Ramírez, Phys. Rev. Lett. **99**, 205504 (2007).

30. H. Maeter, H. Luetkens, Yu. G. Pashkevich, A. Kwadrin, R. Khasanov, A. Amato, A. A. Gusev, K. V. Lamonova, D. A. Chervinskii, R. Klingeler, C. Hess, G. Behr, B. Büchner, and H.-H. Klauss, Phys. Rev. B **80**, 094524 (2009).

31. E. V. Sheely, L. W. Burggraf, P. E. Adamson, X. F. Duan, and M. W. Schmidt, J. Phys.: Conf. Ser. **225**, 012049 (2010).

32. Y. K. Chen, D. G. Fleming, and Y. A. Wang, J. Phys. Chem. A **115**, 2765 (2011).

33. E. L. Silva, A. G. Marinopoulos, R. C. Vilão, R. B. L. Vieira, H. V. Alberto, J. Piroto Duarte, and J. M. Gil, Phys. Rev. B **85**, 165211 (2012).

34. F. Bernardini, P. Bonfà, S. Massidda, and R. De Renzi, Phys. Rev. B **87**, 115148 (2013).

35. S. J. Blundell, J. S. Möller, T. Lancaster, P. J. Baker, F. L. Pratt, G. Seber, and P. M. Lahti, Phys. Rev. B **88**, 064423 (2013).

36. D. G. Fleming, D. J. Arseneau, M. D. Bridges, Y. K. Chen, and Y. A. Wang, J. Phys. Chem. C **117**, 16523 (2013).

37. K. Yamada, Y. Kawashima and M. Tachikawa, J. Chem. Theory Comput. **10**, 2005 (2014)

38. P. Bonfà, F. Sartori and R. De Renzi, J. Phys. Chem. C **119**, 4278 (2015).

39. F. R. Foronda, F. Lang, J. S. Möller, T. Lancaster, A. T. Boothroyd, F. L. Pratt, S. R. Giblin, D. Prabhakaran, and S. J. Blundell, Phys. Rev. Lett. **114**, 017602 (2015).

40. Y. Oba, T. Kawatsu and M. Tachikawa, J. Chem. Phys. **145**, 064301 (2016)

41. R. B. L. Vieira, R. C. Vilão, A. G. Marinopoulos, P. M. Gordo, J. A. Paixão, H. V. Alberto, J. M. Gil, A. Weidinger, R. L. Lichti, B. Baker, P. W. Mengyan, and J. S. Lord, Phys. Rev. B **94**, 115207 (2016).

42. P. Bonfà and R. De Renzi, J. Phys. Soc. Jap. **85**, 091014 (2016).

43. A. Szabo and N. S. Ostlund, *Modern Quantum Chemistry: Introduction to Advanced Electronic Structure Theory* (Dover Publications Inc., New York, 1996).

44. T. Helgaker, P. Jørgenson, and J. Olsen, *Molecular Electronic-Structure Theory* (John Wiley & Sons, New York, 2000).

45. M. Tachikawa, K. Mori, H. Nakai and K. Iguchi, Chem. Phys. Lett. **290**, 437 (1998).

46. S. P. Webb, T. Iordanov and S. Hammes-Schiffer, J. Chem. Phys. **117**, 4106 (2002).

47. H. Nakai, Int. J. Quantum Chem. **107**, 2849 (2007).

48. T. Ishimoto, M. Tachikawa and U. Nagashima, Int. J. Quantum Chem. **109**, 2677 (2009).

49. R. Flores-Moreno, E. Posada, F. Moncada, J. Romero, J. Charry, M. Díaz-Tinoco, S. A. González, N. F. Aguirre and A. Reyes, Int. J. Quantum Chem. **114**, 50 (2014).

50. M. Goli and Sh. Shahbazian, Phys. Chem. Chem. Phys. **16**, 6602 (2014).

51. M. Goli and Sh. Shahbazian, Phys. Chem. Chem. Phys. **17**, 245 (2015).

52. M. Goli and Sh. Shahbazian, Phys. Chem. Chem. Phys. **17**, 7023 (2015).

53. M. Goli and Sh. Shahbazian, Chem. Eur. J. **22**, 2525 (2016).

54. C. Swalina, M. V. Pak and S. Hammes-Schiffer, Chem. Phys. Lett. **404**, 394 (2005).

55. M. Gharabaghi and Sh. Shahbazian, Phys. Lett. A **380**, 3983 (2016).

56. M. Ryaka, M. Goli and Sh. Shahbazian, Phys. Chem. Chem. Phys. **20**, 4466 (2018).

57. M. Ryaka, M. Goli, and Sh. Shahbazian, Phys. Chem. Chem. Phys. **20**, 8802 (2018).

58. A. Chakraborty, M. V. Pak and S. Hammes-Schiffer, J. Chem. Phys. **129**, 014101 (2008); Erratum: 134, 079902 (2011).

59. M. V. Pak, A. Chakraborty and S. Hammes-Schiffer, J. Phys. Chem. A **111**, 4522 (2007).

60. A. Chakraborty, M. V. Pak and S. Hammes-Schiffer, Phys. Rev. Lett. **101**, 153001 (2008); Erratum: **106**, 169902 (2011).

61. A. Chakraborty, M. V. Pak and S. Hammes-Schiffer, J. Chem. Phys. **131**, 124115 (2009).

62. A. Sirjoosingh, M. V. Pak and S. Hammes-Schiffer, J. Chem. Theory Comput. **7**, 2689 (2011).

63. A. Sirjoosingh, M. V. Pak and S. Hammes-Schiffer, J. Chem. Phys. **136**, 174114 (2012).

64. T. Culpitt, K. R. Brorsen, M. V. Pak and S. Hammes-Schiffer, J. Chem. Phys. **145**, 044106 (2016).

65. T. Udagawa, T. Tsuneda and M. Tachikawa, Phys. Rev. A **89**, 052519 (2014).

66. Y. Imamura, H. Kiryu and H. Nakai, J. Comput. Chem. **29**, 735 (2008).

67. Y. Imamura, Y. Tsukamoto, H. Kiryu and H. Nakai, Bull. Chem. Soc. Jpn. **82**, 1133 (2009).

68. T. Kreibich, van Leeuwen R. and E. K. U. Gross, Phys. Rev. Lett. **86**, 2984 (2001).

69. C. Swalina, M.V. Pak, A. Chakraborty and S. Hammes-Schiffer, J. Phys. Chem. A **110**, 9983 (2006).

70. C. Ko, M. V. Pak, C. Swalina and S. Hammes-Schiffer, J. Chem. Phys. **135**, 054106 (2011).

71. C. Swalina, M. V. Pak and S. Hammes-Schiffer, J. Chem. Phys. **136**, 164105 (2012).

72. A. Sirjoosingh, M. V. Pak, C. Swalina and S. Hammes-Schiffer, J. Chem. Phys. **139**, 034102 (2013).

73. A. Sirjoosingh, M. V. Pak, C. Swalina and S. Hammes-Schiffer, J. Chem. Phys. **139**, 034103 (2013).

74. A. Sirjoosingh, M. V. Pak, K. R. Brorsen and S. Hammes-Schiffer, J. Chem. Phys. **142**, 214107 (2015).

75. K. R. Brorsen, A. Sirjoosingh, M. V. Pak and S. Hammes-Schiffer, J. Chem. Phys. **142**, 214108 (2015).

76. Y. Yang, K. R. Brorsen, T. Culpitt, M. V. Pak and S. Hammes-Schiffer, J. Chem. Phys. **147**, 114113 (2017).

77. K. R. Brorsen, Y. Yang, and S. Hammes-Schiffer, J. Phys. Chem. Lett. **8**, 3488 (2017).

78. M. Hoshino, H. Nishizawa and H. Nakai, J. Chem. Phys. **135**, 024111 (2011).

79. T. Ishimoto, M. Tachikawa and U. Nagashima, J. Chem. Phys. **125**, 144103 (2006).

80. A. Kerridge, A. H. Harker and A. M. Stoneham, J. Phys.: Condens. Matter **16**, 8743 (2004).

81. P. Cassam-Chenaï, B. Suo and W. Liu, Phys. Rev. A **92**, 012502 (2015).

82. P. Cassam-Chenaï, B. Suo and W. Liu, Theor. Chem. Acc. **136**, 52 (2017).

83. C. Møller and M. S. Plesset, Phys. Rev. **46**, 618 (1934).

84. J. A. Pople, R. Krishnan, H. B. Schlegel and J. S. Binkley, Int. J. Quantum Chem. **14**, 545 (1978).

85. R. Bartlett and G. D. Purvis, Int. J. Quantum Chem. **14**, 561 (1978).

86. J. Paldus and X. Li, Adv. Chem. Phys. **110**, 1 (1999).

87. R. Bartlett and M. Musiał, Rev. Mod. Phys. **79**, 291 (2007).

88. W. Kutzelnigg, Int. J. Quantum Chem. **109**, 3858 (2009).

89. R. Bartlett, Mol. Phys. **108**, 2905 (2010).

90. C. Swalina, M. V. Pak and S. Hammes-Schiffer, Chem. Phys. Lett. **404**, 394 (2005).

91. M. Hoshino and H. Nakai, J. Chem. Phys. **124**, 194110 (2006).

92. T. Ishimoto, M. Tachikawa and U. Nagashima, J. Chem. Phys. **125**, 144103 (2006).

93. K. Miyamoto, M. Hoshino and H. Nakai, J. Chem. Theory Comput. **2**, 1544 (2006).

94. H. J. Monkhorst, Phys. Rev. A **36**, 1544 (1987).

95. H. Nakai and K. Sodeyama, J. Chem. Phys. **118**, 1119 (2003).

96. H. Nishizawa, Y. Imamura, Y. Ikabata and H. Nakai, Chem. Phys. Lett. **533**, 100 (2012).

97. B. H. Ellis, S. Aggarwal and A. Chakraborty, J. Chem. Theory Comput. **12**, 188 (2016).

98. C. Swalina and S. Hammes-Schiffer, J. Phys. Chem. A **109**, 10410 (2005).

99. P. E. Adamson, X. F. Duan, L. W. Burggraf, M. V. Pak, C. Swalina and S. Hammes-Schiffer, J. Phys. Chem. A **112**, 1346 (2008).

100. Y. Kikuta, T. Ishimoto and U. Nagashima, Bull. Chem. Soc. Jpn. **81**, 820 (2008).

101. D. V. Moreno, S. A. González and A. Reyes, J. Phys. Chem. A **114**, 9231 (2010).

102. S. A. González and A. Reyes, Int. J. Quantum Chem. **110**, 689 (2010).

103. Y. Ikabata, Y. Imamura and H. Nakai, J. Phys. Chem. A **115**, 1433 (2011).

104. D. V. Moreno, S. A. González and A. Reyes, J. Chem. Phys. **134**, 024115 (2011).

105. F. Moncada, L. S. Uribe, J. Romero and A. Reyes, Int. J. Quantum Chem. **113**, 1556 (2013).

106. T. Udagawa and M. Tachikawa, J. Comput. Chem. **35**, 271 (2014).

107. T. Udagawa, T. Ishimoto and M. Tachikawa, Chem. Phys. **441**, 101 (2014).

108. T. Udagawa and M. Tachikawa, J. Comput. Chem. **36**, 1647 (2015).

109. M. Hashimoto, T. Ishimoto, M. Tachikawa and T. Udagawa, Int. J. Quantum Chem. **116**, 961 (2016).

110. J. Olsen, O. Christiansen, H. Koch and P. Jørgenson, J. Chem. Phys. **105**, 5082 (1996).

111. D. Cremer and Z. He, J. Phys. Chem. **100**, 6173 (1996).

112. G. D. Purvis III and R. J. Bartlett, J. Chem. Phys. **76**, 1910 (1982).

113. G. E. Scuseria, C. L. Janssen and H. F. Schaefer III, J. Chem. Phys. **89**, 7382 (1988).

114. J. A. Pople, M. Head-Gordon and K. Raghavachari, J. Chem. Phys. **87**, 5968 (1987).

115. T. Udagawa and M. Tachikawa, J. Chem. Phys. **145**, 164310 (2016).

116. T. Udagawa, K. Suzuki and M. Tachikawa, Proc. Comp. Sci. **108C**, 2275 (2017).

117. T.H. Dunning, Jr., J. Chem. Phys. **90**, 1007 (1989).

118. D.E. Woon and T.H. Dunning, Jr., J. Chem. Phys. **98**, 1358 (1993).

119. A. D. Becke, Phys. Rev. A **38**, 3098 (1988).

120. C. Lee, W. Yang, and R. G. Parr, Phys. Rev. B **37**, 785 (1988).

121. A. D. Becke, J. Chem. Phys. **98**, 5648 (1993).

122. W. J. Lauderdale, J. F. Stanton, J. Gauss, J. D. Watts and R. J. Bartlett, Chem. Phys. Lett. **187**, 21 (1991).

123. P. J. Knowles, J. S. Andrews, R. D. Amos, N. C. Handy, J. A. Pople, Chem. Phys. Lett. **187**, 130 (1991).

124. P. Piecuch and M. Włoch, J. Chem. Phys. **123**, 224105 (2005).

125. M. Włoch, J. R. Gour and P. Piecuch, J. Phys. Chem. A **111**, 11359 (2007).

126. M. W. Schmidt, K. K. Baldridge, J. A. Boatz, S. T. Elbert, M. S. Gordon, J. H. Jensen, S. Koseki, N. Matsunaga, K. A. Nguyen, S. Su, T. L. Windus, M. Dupuis and J. A. Montgomery, J. Comput. Chem. **14**, 1347 (1993).

127. M. S. Gordon and M. W. Schmidt, *Theory and Applications of Computational Chemistry: the first forty years*, pp. 1167-1189, Eds. C. E. Dykstra, G. Frenking, K. S. Kim, G. E. Scuseria (Elsevier, Amsterdam, 2005).

128. R. West and P. W. Percival, Dalton Trans. **39**, 9209 (2010).

129. I. McKenzie, J.-C. Brodovitch, P. W. Percival, T. Ramnial and J. A. C. Clyburne, J. Am. Chem. Soc. **125**, 11565 (2003).

130. B. M. McCollum, T. Abe, J.-C. Brodovitch, J. A. C. Clyburne, T. Iwamoto, M. Kira, P. W. Percival and R. West, Angew. Chem. Int. Ed. **47**, 9772 (2008).

131. B. M. McCollum, J.-C. Brodovitch, J. A. C. Clyburne, P. W. Percival and R. West, Physica B **404**, 940 (2009).

132. B. M. McCollum, J.-C. Brodovitch, J. A. C. Clyburne, A. Mitra, P. W. Percival, A. Tomasik and R. West, Chem. Eur. J. **15**, 8409 (2009).

133. P. W. Percival, J.-C. Brodovitch, M. Mozafari, A. Mitra, R. West, R. S. Ghadwal, R. Azhakar and H. W. Roesky, Chem. Eur. J. **17**, 11970 (2011).

134. P. W. Percival, B. M. McCollum, J.-C. Brodovitch, M. Driess, A. Mitra, M. Mozafari, R. West, Y. Xiong and Sh. Yai, Organometallics **31**, 2709 (2012).

135. R. West, K. Samedov, A. Mitra, P. W. Percival, J.-C. Brodovitch, G. Langille, B. M. McCollum, T. Iwamoto, Sh. Ishida, C. Jones and J. Li, Can. J. Chem. **92**, 508 (2014).

136. K. Samedov, R. West, P. W. Percival, J.-C. Brodovitch, L. Chandrasena, M. Mozafari, R. Tacke, K. Junold, C. Kobelt, P. P. Samuel, R. Azhakar, K. Ch. Mondal, H. W. Roesky, M. Driess and W. Wang, Organometallics **34**, 3532 (2015).

137. W. Herrmann and Ch. Köcher, Angew. Chem. Int. Ed. **36**, 2162 (1997).

138. A. J. Arduengo, III Acc. Chem. Res. **32**, 913 (1999).

139. M. Haaf, T. A. Schmedake and R. West, Acc. Chem. Res. **33**, 704 (2000).

140. N. J. Hill and R. West, J. Organomet. Chem. **689**, 4165 (2004).

141. O. Kühl, Coord. Chem. Rev. **248**, 411 (2004).

142. M. N. Hopkinson, Ch. Richter, M. Schedler and F. Glorius, Nature **510**, 485 (2014).

143. W. Kirmse, Angew. Chem. Int. Ed. **49**, 8798 (2010).

144. F. E. Hahn and M. C. Jahnke, Angew. Chem. Int. Ed. **47**, 3122 (2008).

145. Y. Mizuhata, T. Sasamori and N. Tokitoh, Chem. Rev. **109**, 3479 (2009).

146. V. Ya. Lee and A. Sekiguchi, *Organometallic Compounds of Low-Coordinate Si, Ge, Sn and Pb: From Phantom Species to Stable Compounds* (John Wiley & Sons Ltd, United Kingdom, 2010).

147. M. Asay, C. Jones and M. Driess, Chem. Rev. **111**, 354 (2011).

148. G. I. Borodkin and V. G. Shubin, Russ. Chem. Rev. **77**, 395 (2008).

149. Y. Tulchinsky, M. A. Iron, M. Botoshansky and M. Gandelman, Nature Chem. **3**, 525 (2011).

150. Z. Yacob and J. Liebscher, Top. Heterocycl. Chem. **40**, 167 (2014).

151. G. A. McGibbon, Ch. Heinemann, D. J. Lavorato and H. Schwarz, Angew. Chem. Int. Ed. **36**, 1478 (1997).

152. G. Maier and J. Endres, Eur. J. Org. Chem. 1517 (1998).

153. L. Pause, M. Robert, J. Heinicke and O. Kühl, J. Chem. Soc. Perkin Trans. 2, 1383 (2001).

154. C. D. Martin, M. Soleilhavoup and G. Bertrand, Chem. Sci. **4**, 3020 (2013).

155. Ch. Chatgilialoglu, Chem. Rev. **95**, 1229 (1995).

156. F. M. Bickelhaupt, T. Ziegler, P. v. R. Schleyer, Organometallics **15**, 1477 (1996).

157. R. F. W. Bader, *Atoms in Molecules: A Quantum Theory* (Oxford University Press, Oxford, 1990).

158. M. Goli and Sh. Shahbazian, Theor. Chem. Acc. **132**, 1365 (2013).

159. R. M. Macrae, Physica B **374-375**, 307 (2006).

160. J. A. Wright, J. N. T. Peck, S. P. Cottrell, A. Jablonskytė, V. S. Oganesyan, C. J. Pickett and U. A. Jayasooriya, Angew. Chem. Int. Ed. **55**, 14580 (2016).

161. I. McKenzie, Can. J. Chem. **96**, 358 (2018).

# Supporting Information

**Developing effective electronic-only coupled-cluster and Møller-Plesset perturbation theories for the muonic molecules**

Mohammad Goli[1,*] and Shant Shahbazian[2,*]

[1] *School of Nano Science, Institute for Research in Fundamental Sciences (IPM), Tehran 19395-5531, Iran*

[2] *Department of Physics, Shahid Beheshti University, G. C., Evin, Tehran, Iran, 19839, P.O. Box 19395-4716.*

E-mails:

Mohammad Goli: m_goli @ipm.ir

Shant Shahbazian: sh_shahbazian@sbu.ac.ir

[*] Corresponding authors

# Table of contents

Table S1- The distance between hydrogen and central atoms computed at both HF and post-HF levels (in Angstroms) for hydrides.

|  | HF | MP2 | CCSD | CCSD(T) |
|---|---|---|---|---|
| **LiH** | 1.607 | 1.604 | 1.610 | 1.610 |
| **BeH$_2$** | 1.331 | 1.329 | 1.333 | 1.333 |
| **BH$_3$** | 1.188 | 1.187 | 1.191 | 1.191 |
| **CH$_4$** | 1.082 | 1.086 | 1.088 | 1.090 |
| **NH$_3$** | 0.999 | 1.012 | 1.012 | 1.015 |
| **H$_2$O** | 0.941 | 0.961 | 0.959 | 0.962 |
| **FH** | 0.899 | 0.921 | 0.918 | 0.921 |
| **NaH** | 1.919 | 1.918 | 1.926 | 1.926 |
| **MgH$_2$** | 1.708 | 1.706 | 1.712 | 1.713 |
| **AlH$_3$** | 1.581 | 1.580 | 1.585 | 1.585 |
| **SiH$_4$** | 1.478 | 1.478 | 1.482 | 1.483 |
| **PH$_3$** | 1.408 | 1.412 | 1.417 | 1.419 |
| **H$_2$S** | 1.330 | 1.336 | 1.340 | 1.342 |
| **HCl** | 1.268 | 1.275 | 1.277 | 1.279 |

Table S2- The difference (in Angstroms) between the muonic bond lenghts (in Table 1) and the analog bond lengths in hydrides (in Table S1).

|  | EHF-HF | EMP2-MP2 | ECCSD-CCSD | ECCSD(T)-CCSD(T) |
|---|---|---|---|---|
| LiMu-LiH | 0.089 | 0.085 | 0.083 | 0.083 |
| BeHMu-BeH$_2$ | 0.082 | 0.080 | 0.078 | 0.078 |
| BH$_2$Mu-BeH$_3$ | 0.077 | 0.070 | 0.071 | 0.070 |
| CH$_3$Mu-CH$_4$ | 0.069 | 0.062 | 0.061 | 0.060 |
| NH$_2$Mu-NH$_3$ | 0.061 | 0.046 | 0.046 | 0.044 |
| OHMu-H$_2$O | 0.056 | 0.036 | 0.038 | 0.036 |
| FMu-FH | 0.055 | 0.033 | 0.036 | 0.034 |
| NaMu-NaH | 0.079 | 0.075 | 0.072 | 0.072 |
| MgHMu-MgH$_2$ | 0.083 | 0.078 | 0.074 | 0.074 |
| AlH$_2$Mu-AlH$_3$ | 0.083 | 0.078 | 0.075 | 0.074 |
| SiH$_3$Mu-SiH$_4$ | 0.081 | 0.075 | 0.073 | 0.072 |
| PH$_2$Mu-PH$_3$ | 0.081 | 0.072 | 0.068 | 0.067 |
| SHMu-H$_2$S | 0.076 | 0.064 | 0.062 | 0.059 |
| MuCl-HCl | 0.074 | 0.061 | 0.059 | 0.057 |

Table S3- The bond lengths of the vicinal clamped hydrogen to the central atoms computed at both HF and post-HF levels (in Angstroms) for the muonic species.

|  | EHF | EMP2 | ECCSD | ECCSD(T) |
|---|---|---|---|---|
| **LiMu** | -- | -- | -- | -- |
| **BeHMu** | 1.332 | 1.330 | 1.334 | 1.334 |
| **BH$_2$Mu** | 1.187 | 1.185 | 1.191 | 1.192 |
| **CH$_3$Mu** | 1.082 | 1.086 | 1.088 | 1.089 |
| **NH$_2$Mu** | 0.999 | 1.012 | 1.012 | 1.015 |
| **OHMu** | 0.941 | 0.962 | 0.959 | 0.962 |
| **FMu** | -- | -- | -- | -- |
| **NaMu** | -- | -- | -- | -- |
| **MgHMu** | 1.708 | 1.707 | 1.713 | 1.714 |
| **AlH$_2$Mu** | 1.582 | 1.581 | 1.585 | 1.586 |
| **SiH$_3$Mu** | 1.478 | 1.478 | 1.482 | 1.484 |
| **PH$_2$Mu** | 1.408 | 1.413 | 1.417 | 1.420 |
| **SHMu** | 1.330 | 1.337 | 1.340 | 1.343 |
| **MuCl** | -- | -- | -- | -- |

Table S4- The total energies at corresponding optimized geometries (in Hartrees) for the muonic species and the analog hydrides.[*]

|  | EHF | EMP2 | ECCSD | ECCSD(T) |
|---|---|---|---|---|
| **LiMu** | -7.8917 | -7.9185 | -7.9265 | -7.9265 |
| **BeHMu** | -15.6687 | -15.7300 | -15.7452 | -15.7459 |
| **BH$_2$Mu** | -26.2914 | -26.4070 | -26.4284 | -26.4312 |
| **CH$_3$Mu** | -40.1034 | -40.3067 | -40.3258 | -40.3326 |
| **NH$_2$Mu** | -56.1117 | -56.3551 | -56.3655 | -56.3742 |
| **OHMu** | -75.9552 | -76.2281 | -76.2314 | -76.2405 |
| **FMu** | -99.9606 | -100.2464 | -100.2464 | -100.2543 |
| **NaMu** | -162.2992 | -162.3259 | -162.3344 | -162.3344 |
| **MgHMu** | -200.6408 | -200.6986 | -200.7143 | -200.7148 |
| **AlH$_2$Mu** | -243.5433 | -243.6393 | -243.6623 | -243.6641 |
| **SiH$_3$Mu** | -291.1569 | -291.3041 | -291.3332 | -291.3375 |
| **PH$_2$Mu** | -342.3847 | -342.5597 | -342.5869 | -342.5943 |
| **SHMu** | -398.6128 | -398.8097 | -398.8321 | -398.8410 |
| **MuCl** | -460.0102 | -460.2200 | -460.2384 | -460.2472 |
|  | **HF** | **MP2** | **CCSD** | **CCSD(T)** |
| **LiH** | -7.9870 | -8.0153 | -8.0234 | -8.0234 |
| **BeH$_2$** | -15.7721 | -15.8341 | -15.8492 | -15.8498 |
| **BH$_3$** | -26.4004 | -26.5157 | -26.5370 | -26.5397 |
| **CH$_4$** | -40.2139 | -40.4157 | -40.4354 | -40.4420 |
| **NH$_3$** | -56.2211 | -56.4620 | -56.4735 | -56.4819 |
| **H$_2$O** | -76.0617 | -76.3312 | -76.3358 | -76.3445 |
| **FH** | -100.0623 | -100.3446 | -100.3458 | -100.3534 |
| **NaH** | -162.3912 | -162.4197 | -162.4286 | -162.4286 |
| **MgH$_2$** | -200.7384 | -200.7972 | -200.8132 | -200.8136 |
| **AlH$_3$** | -243.6453 | -243.7418 | -243.7649 | -243.7666 |
| **SiH$_4$** | -291.2616 | -291.4085 | -291.4378 | -291.4419 |
| **PH$_3$** | -342.4885 | -342.6625 | -342.6903 | -342.6974 |
| **H$_2$S** | -398.7143 | -398.9101 | -398.9334 | -398.9420 |
| **HCl** | -460.1080 | -460.3171 | -460.3366 | -460.3452 |

[*] Note that the correlation energies reported in Table 2 of the main text is *not* the difference between the EHF/HF and post-EHF/HF total energies reported in this table, which are computed at different optimized geometries. The correlation energies in Table 2 are the difference between post-EHF/HF and the EHF/HF energies at the corresponding post-EHF/HF optimized geometries.

Table S5- The total energies (in Hartrees) for the hydrogenated and the muoniated adducts (for details see the main text).

|            | B3LYP       | MP2         | CCSD        |
|------------|-------------|-------------|-------------|
| H-cC-NHC   | -226.84088  | -226.31437  | -226.33621  |
| H-aC-NHC   | -226.81870  | -226.28751  | -226.31704  |
| H-N-TAC    | -243.22921  | -242.69246  | -242.70513  |
| H-C-TAC    | -243.22332  | -242.67744  | -242.70177  |
| H-Si-NHSi  | -478.31277  | -477.40208  | -477.42712  |
| H-C-NHSi   | -478.30383  | -477.39391  | -477.42554  |
| H-Ge-NHGe  | -2265.87997 | -2263.85457 | -2263.87800 |
| H-C-NHGe   | -2265.87359 | -2263.84930 | -2263.87687 |
|            | EB3LYP      | EMP2        | ECCSD       |
| Mu-cC-NHC  | -226.73044  | -226.20609  | -226.22674  |
| Mu-aC-NHC  | -226.70817  | -226.17889  | -226.20742  |
| Mu-N-TAC   | -243.12379  | -242.59041  | -242.60056  |
| Mu-C-TAC   | -243.11531  | -242.57072  | -242.59446  |
| Mu-Si-NHSi | -478.20860  | -477.29928  | -477.32386  |
| Mu-C-NHSi  | -478.19280  | -477.28469  | -477.31536  |
| Mu-Ge-NHGe | -2265.77793 | -2263.75430 | -2263.77682 |
| Mu-C-NHGe  | -2265.76280 | -2263.74025 | -2263.76676 |

## parent N-Heterocyclic carbene (NHC)

| atom | X | Y | Z |
|---|---|---|---|
| C | 0.070944 | 0.000000 | -0.001735 |
| H | -0.075400 | 0.000000 | 1.062793 |
| C | 1.204366 | 0.000000 | -0.735965 |
| H | 2.235705 | 0.000000 | -0.434344 |
| N | 0.790425 | 0.000000 | -2.062069 |
| H | 1.413779 | 0.000000 | -2.848918 |
| N | -0.970072 | 0.000000 | -0.921622 |
| H | -1.942998 | 0.000000 | -0.674405 |
| C | -0.562129 | 0.000000 | -2.220938 |

## parent N-Heterocyclic silylene (NHSi)

| atom | X | Y | Z |
|---|---|---|---|
| C | 0.048524 | 0.000000 | -0.035639 |
| H | -0.013346 | 0.000000 | 1.040245 |
| C | 1.182589 | 0.000000 | -0.770287 |
| H | 2.189757 | 0.000000 | -0.386927 |
| N | 0.921378 | 0.000000 | -2.132281 |
| H | 1.700681 | 0.000000 | -2.768784 |
| N | -1.087680 | 0.000000 | -0.830813 |
| H | -1.987148 | 0.000000 | -0.379807 |
| Si | -0.790134 | 0.000000 | -2.572907 |

## parent N-Heterocyclic germylene (NHGe)

| atom | X | Y | Z |
|---|---|---|---|
| C | 0.040559 | 0.000000 | -0.043083 |
| H | 0.002864 | 0.000000 | 1.035062 |
| C | 1.179051 | 0.000000 | -0.780599 |
| H | 2.178401 | 0.000000 | -0.374251 |
| N | 0.958486 | 0.000000 | -2.139720 |
| H | 1.758559 | 0.000000 | -2.750829 |
| N | -1.109639 | 0.000000 | -0.799989 |
| H | -1.994423 | 0.000000 | -0.319647 |
| Ge | -0.849238 | 0.000000 | -2.664145 |

## triazolium cation (TAC)

| atom | X | Y | Z |
|------|-----------|----------|-----------|
| C | 0.064639 | 0.000000 | 0.004819 |
| H | -0.104264 | 0.000000 | 1.066989 |
| C | 1.212923 | 0.000000 | -0.739038 |
| H | 2.251332 | 0.000000 | -0.458973 |
| N | 0.786060 | 0.000000 | -2.027153 |
| H | 1.342821 | 0.000000 | -2.874602 |
| N | -0.936424 | 0.000000 | -0.911334 |
| H | -1.937429 | 0.000000 | -0.749660 |
| N | -0.515037 | 0.000000 | -2.148248 |

## H-cC-NHC

| atom | X | Y | Z |
|------|-----------|-----------|-----------|
| C | 0.671566 | 0.978634 | -0.005403 |
| H | 1.353582 | 1.809759 | 0.009675 |
| C | -0.671567 | 0.978633 | -0.005404 |
| H | -1.353584 | 1.809757 | 0.009675 |
| N | -1.117536 | -0.343232 | 0.071695 |
| H | -2.006349 | -0.598628 | -0.326600 |
| N | 1.117536 | -0.343230 | 0.071695 |
| H | 2.006350 | -0.598625 | -0.326600 |
| C | 0.000001 | -1.201322 | -0.119089 |
| H | 0.000001 | -2.153033 | 0.406653 |

## H-aC-NHC

| atom | X | Y | Z |
|------|-----------|-----------|-----------|
| C | 0.486518 | -1.044784 | -0.197995 |
| H | 0.891583 | -1.967560 | -0.573209 |
| N | -0.875293 | -0.771218 | -0.178251 |
| H | -1.585672 | -1.475495 | -0.258924 |
| N | -0.004509 | 1.141233 | 0.106789 |
| H | 0.074829 | 2.131291 | 0.246710 |
| C | -1.201036 | 0.539529 | 0.030948 |
| C | 1.172540 | 0.271016 | -0.069760 |
| H | 1.854650 | 0.335280 | 0.787897 |
| H | 1.752871 | 0.555374 | -0.960374 |

**H-Si-NHSi**

| atom | X | Y | Z |
|------|-----------|-----------|-----------|
| C  | -1.284279 |  0.679186 |  0.010600 |
| H  | -2.163743 |  1.299937 | -0.047200 |
| C  | -1.284279 | -0.679188 |  0.010579 |
| H  | -2.163742 | -1.299938 | -0.047231 |
| N  | -0.020097 | -1.228519 |  0.055331 |
| H  |  0.070784 | -2.225552 | -0.033477 |
| N  | -0.020096 |  1.228518 |  0.055323 |
| H  |  0.070787 |  2.225551 | -0.033494 |
| Si |  1.271695 | -0.000002 | -0.132177 |
| H  |  2.042087 |  0.000006 |  1.174246 |

**H-C-NHSi**

| atom | X | Y | Z |
|------|-----------|-----------|-----------|
| C  |  1.143943 | -0.861844 |  0.000000 |
| H  |  1.935578 | -1.592377 |  0.000000 |
| N  | -0.180305 | -1.220611 |  0.000000 |
| H  | -0.404105 | -2.202422 |  0.000000 |
| N  | -0.000019 |  1.186920 |  0.000000 |
| H  | -0.061640 |  2.190918 |  0.000000 |
| Si | -1.358639 |  0.101079 |  0.000000 |
| C  |  1.350083 |  0.616694 |  0.000000 |
| H  |  1.926760 |  0.949178 |  0.878447 |
| H  |  1.926760 |  0.949178 | -0.878447 |

**H-Ge-NHGe**

| atom | X | Y | Z |
|------|-----------|-----------|-----------|
| C  | -1.712638 |  0.691910 | -0.008819 |
| H  | -2.617521 |  1.277233 | -0.092619 |
| C  | -1.712636 | -0.691912 | -0.008823 |
| H  | -2.617515 | -1.277240 | -0.092626 |
| N  | -0.501294 | -1.270568 |  0.071946 |
| H  | -0.452687 | -2.273328 | -0.010462 |
| N  | -0.501297 |  1.270569 |  0.071943 |
| H  | -0.452691 |  2.273332 | -0.010465 |
| Ge |  1.010729 |  0.000000 | -0.068399 |
| H  |  1.366874 |  0.000003 |  1.493561 |

### H-C-NHGe

| atom | X | Y | Z |
|------|-----------|-----------|-----------|
| C  | -0.775185 | -1.638381 |  0.000000 |
| H  | -1.425510 | -2.499697 |  0.000000 |
| N  | -1.263398 | -0.379850 |  0.000000 |
| H  | -2.265202 | -0.265742 |  0.000000 |
| N  |  1.218351 | -0.370197 |  0.000000 |
| H  |  2.219021 | -0.264855 |  0.000000 |
| Ge |  0.000850 |  1.036347 |  0.000000 |
| C  |  0.715228 | -1.738472 |  0.000000 |
| H  |  1.071160 | -2.306866 |  0.877407 |
| H  |  1.071160 | -2.306866 | -0.877407 |

### H-N-TAC

| atom | X | Y | Z |
|------|-----------|-----------|-----------|
| C | -0.680487 |  0.992095 |  0.008361 |
| H | -1.375698 |  1.812736 | -0.009311 |
| C |  0.680492 |  0.992092 |  0.008362 |
| H |  1.375708 |  1.812729 | -0.009308 |
| N |  1.097529 | -0.304909 |  0.048048 |
| H |  1.990096 | -0.648849 | -0.283385 |
| N | -1.097530 | -0.304904 |  0.048049 |
| H | -1.990100 | -0.648839 | -0.283383 |
| N | -0.000003 | -1.151603 | -0.107844 |
| H | -0.000005 | -1.916289 |  0.569785 |

### H-C-TAC

| atom | X | Y | Z |
|------|-----------|-----------|-----------|
| C |  0.523599 | -1.054702 |  0.000008 |
| H |  0.944408 | -2.045400 |  0.000328 |
| N | -0.778609 | -0.819784 | -0.000175 |
| H | -1.527564 | -1.504503 | -0.000051 |
| N | -0.000131 |  1.112835 |  0.000116 |
| H | -0.041099 |  2.122330 |  0.000408 |
| N | -1.132251 |  0.456859 | -0.000161 |
| C |  1.199589 |  0.270335 | -0.000007 |
| H |  1.827688 |  0.444663 |  0.884052 |
| H |  1.827179 |  0.444014 | -0.884519 |

**Mu-cC-NHC**

| atom | X | Y | Z |
|------|-----------|-----------|-----------|
| C    | 0.671609  | 0.985203  | -0.006768 |
| H    | 1.353921  | 1.816064  | 0.013696  |
| C    | -0.671495 | 0.985210  | -0.007292 |
| H    | -1.353850 | 1.816065  | 0.013113  |
| N    | -1.115712 | -0.337112 | 0.063533  |
| H    | -2.005644 | -0.594382 | -0.331208 |
| N    | 1.115895  | -0.337254 | 0.064172  |
| H    | 2.005225  | -0.594116 | -0.332052 |
| C    | 0.000041  | -1.199165 | -0.117447 |
| Bq   | 0.000011  | -2.201800 | 0.426548  |

**Mu-aC-NHC**

| atom | X | Y | Z |
|------|-----------|-----------|-----------|
| C    | 0.483972  | -1.045086 | -0.182049 |
| H    | 0.890641  | -1.970403 | -0.549491 |
| N    | -0.878317 | -0.769807 | -0.186302 |
| H    | -1.587997 | -1.473553 | -0.277234 |
| N    | -0.007864 | 1.137098  | 0.125869  |
| H    | 0.071330  | 2.128157  | 0.258516  |
| C    | -1.205114 | 0.539353  | 0.027770  |
| C    | 1.167097  | 0.268116  | -0.061458 |
| H    | 1.863361  | 0.335704  | 0.782592  |
| Bq   | 1.769370  | 0.565087  | -1.004384 |

**Mu-Si-NHSi**

| atom | X | Y | Z |
|------|-----------|-----------|-----------|
| C    | -1.289712 | 0.678870  | 0.008953  |
| H    | -2.169042 | 1.300999  | -0.033849 |
| C    | -1.289771 | -0.678978 | 0.008890  |
| H    | -2.169252 | -1.300932 | -0.033626 |
| N    | -0.023904 | -1.226537 | 0.033410  |
| H    | 0.068124  | -2.224311 | -0.045004 |
| N    | -0.023810 | 1.226315  | 0.033545  |
| H    | 0.068428  | 2.224089  | -0.044579 |
| Si   | 1.270107  | -0.000157 | -0.148462 |
| Bq   | 2.077949  | 0.000643  | 1.233223  |

### Mu-C-NHSi

| atom | X | Y | Z |
|------|------|------|------|
| C  |  1.139992 | -0.860289 |  0.059105 |
| H  |  1.932048 | -1.589876 |  0.023282 |
| N  | -0.182138 | -1.221086 | -0.016394 |
| H  | -0.404259 | -2.203137 | -0.031161 |
| N  | -0.003534 |  1.184604 |  0.035850 |
| H  | -0.065399 |  2.188502 |  0.027486 |
| Si | -1.361759 |  0.099034 | -0.028738 |
| C  |  1.344693 |  0.612792 |  0.013394 |
| H  |  1.952455 |  0.969817 |  0.856777 |
| Bq |  1.926316 |  0.936352 | -0.939601 |

### Mu-Ge-NHGe

| atom | X | Y | Z |
|------|------|------|------|
| C  | -1.718225 |  0.690833 | -0.011272 |
| H  | -2.622440 |  1.278367 | -0.083779 |
| C  | -1.718249 | -0.690733 | -0.010428 |
| H  | -2.622495 | -1.278487 | -0.080928 |
| N  | -0.502970 | -1.268065 |  0.055538 |
| H  | -0.452245 | -2.270997 | -0.021511 |
| N  | -0.502846 |  1.267953 |  0.054289 |
| H  | -0.452311 |  2.271192 | -0.018854 |
| Ge |  1.002727 | -0.000138 | -0.085374 |
| Bq |  1.394116 |  0.000073 |  1.553497 |

### Mu-C-NHGe

| atom | X | Y | Z |
|------|------|------|------|
| C  | -0.772806 | -1.633089 |  0.060746 |
| H  | -1.421981 | -2.495241 |  0.083493 |
| N  | -1.262712 | -0.380362 | -0.036875 |
| H  | -2.264817 | -0.268648 | -0.049083 |
| N  |  1.215789 | -0.367904 |  0.085998 |
| H  |  2.215268 | -0.262505 |  0.033980 |
| Ge |  0.001070 |  1.038577 | -0.054143 |
| C  |  0.710182 | -1.731481 |  0.013141 |
| H  |  1.092013 | -2.363305 |  0.828301 |
| Bq |  1.054469 | -2.270622 | -0.965557 |

**Mu-N-TAC**

| atom | X | Y | Z |
|------|-----------|-----------|-----------|
| C  | -0.679135 |  0.998900 |  0.006222 |
| H  | -1.375781 |  1.818190 | -0.010165 |
| C  |  0.679177 |  0.998870 |  0.006124 |
| H  |  1.375732 |  1.818224 | -0.010577 |
| N  |  1.096153 | -0.300885 |  0.048452 |
| H  |  1.985632 | -0.643870 | -0.291657 |
| N  | -1.096255 | -0.300853 |  0.048729 |
| H  | -1.985453 | -0.643815 | -0.292129 |
| N  | -0.000060 | -1.147538 | -0.096295 |
| Bq | -0.000007 | -1.962965 |  0.580669 |

**Mu-C-TAC**

| atom | X | Y | Z |
|------|-----------|-----------|-----------|
| C  |  0.519816 | -1.052118 |  0.012546 |
| H  |  0.940987 | -2.042548 |  0.016560 |
| N  | -0.782688 | -0.820130 | -0.003537 |
| H  | -1.529556 | -1.506783 | -0.013073 |
| N  | -0.004098 |  1.111149 |  0.016891 |
| H  | -0.045134 |  2.120307 |  0.010394 |
| N  | -1.139098 |  0.455349 | -0.004343 |
| C  |  1.192371 |  0.268721 |  0.008073 |
| H  |  1.833921 |  0.444832 |  0.881352 |
| Bq |  1.856288 |  0.447870 | -0.924863 |